\documentclass[a4paper,onecolumn,11pt,unpublished,nopdfoutputerror]{quantumarticle}
\usepackage[utf8]{inputenc}
\usepackage[english]{babel}
\usepackage[T1]{fontenc}
\usepackage{amsmath}
\usepackage{amssymb}
\usepackage{hyperref}
\usepackage[backend=bibtex]{biblatex}
\usepackage[braket]{qcircuit}
\usepackage{mathtools}
\usepackage{pgfplots}
\usepackage{caption}
\usepackage{subcaption}
\usepackage{csquotes}
\usepackage{pgf}
\usepackage{booktabs}
\usepackage[bottom]{footmisc}

\DeclareUnicodeCharacter{2212}{−}
\usepgfplotslibrary{groupplots,dateplot}
\pgfplotsset{compat=newest}
\tikzset{>=latex}
\usetikzlibrary{
    external,
    patterns,
    arrows,
    positioning,
    chains,
    decorations.pathmorphing,
    decorations.markings,
    decorations.pathreplacing,
    shapes,
    shadows.blur,
    shapes.symbols,
    calligraphy,
    fadings,
    calc,
}

\DeclarePairedDelimiter{\ceil}{\lceil}{\rceil}
\DeclareMathOperator{\Ima}{Im}

\begin{document}

\title{Quantum-Enhanced Topological Data Analysis: A Peep from an Implementation Perspective}

\author{Ankit Khandelwal}
\affiliation{TCS Research, Tata Consultancy Services, India}
\email{khandelwal.ankit3@tcs.com}
\author{M Girish Chandra}
\affiliation{TCS Research, Tata Consultancy Services, India}
\email{m.gchandra@tcs.com}

\maketitle

\begin{abstract}
    There is heightened interest in quantum algorithms for Topological Data Analysis (TDA) as it is a powerful tool for data analysis, but it can get highly computationally expensive. Even though there are different propositions and observations for Quantum Topological Data Analysis (QTDA), the necessary details to implement them on software platforms are lacking. Towards closing this gap, the present paper presents an implementation of one such algorithm for calculating Betti numbers.

    The step-by-step instructions for the chosen quantum algorithm and the aspects of how it can be used for machine learning tasks are provided. We provide encouraging results on using Betti numbers for classification and give a preliminary analysis of the effect of the number of shots and precision qubits on the outcome of the quantum algorithm.
\end{abstract}

\section{Introduction}
TDA offers a robust way to extract useful information from noisy, unstructured data by identifying its underlying structure. One essential feature is the number of $k-$dimensional holes and voids in datasets, which is the $k^{th}$ Betti number. Analysis of Betti numbers of data has been growing rapidly in recent years for various applications. Computing the $k^{th}$ Betti number for a given number of data points $n$ involve combinatorially large objects (often referred to as Big Computation problem), which is overwhelming for the most powerful classical computers even for not-so-large $n$. Quantum TDA algorithms, including the possible Noisy Intermediate Scale Quantum (NISQ) amenable versions, are being proposed recently to tackle this problem. Even though possible exponential speed-ups over the existing classical algorithms are spelt out in the earlier works, the complexity advantage aspect is yet to settle down. It appears that there can be a polynomial advantage at least. QTDA algorithms thus can facilitate the extraction of topological features in a computationally efficient way. Machine learning algorithms, both classical and quantum, can then be applied on the features extracted using QTDA techniques.

One of the issues authors of this paper faced while trying to put existing QTDA algorithms into action using the existing Quantum SDKs/software is the lack of explicit implementation steps and the associated circuitry required. To bridge this gap, in this paper, we capture the necessary details required for systematic execution of simulations. This is further augmented with a simple worked-out example. We hope the content can strengthen the understanding of QTDA by facilitating the ``build, run and examine'' approach.

In this paper, we first introduce the concept of TDA and Betti numbers and present the classical algorithm that is used for calculating them. This entails providing the basic idea of the problem and the components that make up the solution. Next, we move to a flavour of QTDA and show how the quantum phase estimation (QPE) algorithm helps estimate the Betti numbers in that approach. The paper also captures novel observations of simulation results obtained by running the QTDA algorithm on randomly generated simplicial complexes and analysing how the number of shots and precision qubits can effect the results. Also, we use the QTDA algorithm to classify time series data and show encouraging results. This paper is the first in a series of planned works on the topic of Quantum-Enhanced TDA.

The paper is organised as follows. In Section~\ref{tda}, we introduce TDA and provide the classical algorithm for calculating Betti numbers. In Section~\ref{qtda}, we present the QTDA algorithm for estimating Betti numbers using (QPE). In Section~\ref{experiment}, we have provided results from applying the QTDA algorithm. In Section~\ref{classification}, we show how the algorithm can be used for classification tasks. Section~\ref{conclusion} concludes the work with some pointers for future work. We also provide a worked-out example in Appendix~\ref{example}.

\begin{figure}
    \centering
    \begin{subfigure}{.2\columnwidth}
        \centering
        \includegraphics[width=\linewidth]{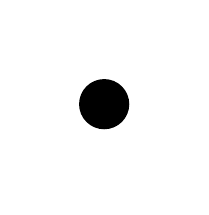}
        \caption{0-simplex}
        \label{fig:0}
    \end{subfigure}
    \hfill
    \begin{subfigure}{.2\columnwidth}
        \centering
        \includegraphics[width=\linewidth]{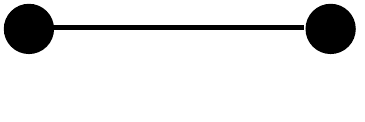}
        \caption{1-simplex}
        \label{fig:1}
    \end{subfigure}
    \hfill
    \begin{subfigure}{.2\columnwidth}
        \centering
        \includegraphics[width=\linewidth]{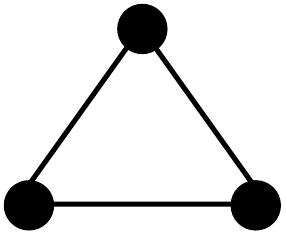}
        \caption{2-simplex}
        \label{fig:2}
    \end{subfigure}
    \hfill
    \begin{subfigure}{.2\columnwidth}
        \centering
        \includegraphics[width=\linewidth]{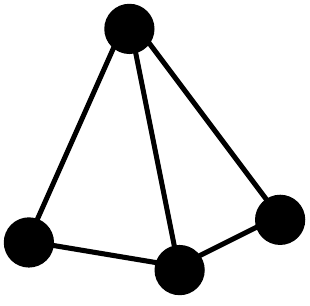}
        \caption{3-simplex}
        \label{fig:3}
    \end{subfigure}
    \caption{First four k-simplicies.}
    \label{fig:k_simplicies}
\end{figure}

\section{Some Essential Points about TDA}
\label{tda}
As mentioned earlier, TDA is getting established as a good candidate for exploratory data analysis by providing useful geometric features. In TDA, the dataset is typically a point cloud, and the aim is to  extract the shape of the underlying data. This is done by constructing a connected object, called a simplicial complex, composed of points, lines, triangles and their higher dimensional counterparts. These are further elaborated in the following.

We start by making a point cloud of the dataset ($\{x_i\}_{i=1}^{n}$) in an $m$ dimensional space where $m$ is the number of features in the dataset, and $n$ is the number of data points (see also Section~\ref{classification}). This space has a distance function $d(x_i,x_j)$ defined on it, usually the Euclidean distance on $\mathbb{R}^m$. Using this distance function, we connect all points within $\epsilon$ (grouping scale) distance from each other with edges. This creates a graph $G_\epsilon=(V,E_\epsilon)$, with vertices $V=\{v_1,v_2,\dots,v_n\}$ and edges $E_\epsilon=\{(i,j) \mid d(x_i,x_j)\leq\epsilon \}$.

A $k$-simplex is a collection of $k+1$ vertices with $k(k+1)/2$ edges in $k$ dimensions. The first few $k$-simplicies are shown in Fig.~\ref{fig:k_simplicies}. A number of these simplicies are present in the graph $G_\epsilon$, which are together called the simplicial complex $\mathcal{K}_\epsilon$.

Let $S_k^\epsilon$ be the set of $k$-simplicies in the complex $\mathcal{K}_\epsilon$ with individual simplicies denoted by $s_k^\epsilon\in S_k^\epsilon$ written as $[j_0,j_1,\dots,j_k]$ where $j_i$ is the $i^{th}$ vertex of $s_k^\epsilon$. Note that the vertices are ordered in ascending fashion in the initial point cloud, and this order is kept throughout. 

The restricted boundary operator $\partial_k^\epsilon$ is defined on the $k$-simplicies as \cite{ubaru2021quantum,towards2022}:
\begin{align}
    \partial_k^\epsilon s_k^\epsilon&=\sum_{t=0}^{k}(-1)^t[v_0,\dots,v_{t-1},v_{t+1},\dots,v_k]\\
    &=\sum_{t=0}^{k}(-1)^t s^\epsilon_{k-1}(t)
\end{align}
where $s_{k-1}^\epsilon(t)$ is defined as the lower simplex defined from $s_k^\epsilon$ by leaving out the vertex $v_t$.

From $\partial_k^\epsilon$ we get the $k$-homology group defined as the quotient space $\mathbb{H}_k^\epsilon$ \cite{berry2022quantifying,ubaru2021quantum,towards2022}:
\begin{equation}
    \mathbb{H}_k^\epsilon=\frac{\ker\partial_k^\epsilon}{\Ima\partial_{k+1}^\epsilon}
\end{equation}
where $\ker A$ and $\Ima A$ are the kernel (null space, set of solutions of $A\Vec{x}=\Vec{0}$) and image (set of all outputs $A\Vec{x}$) of $A$.

The $k^{th}$ Betti number $\beta_k^\epsilon$ is the dimension of $\mathbb{H}_k^\epsilon$ \cite{berry2022quantifying,ubaru2021quantum,towards2022}:
\begin{equation}
    \beta_k^\epsilon=\dim\mathbb{H}_k^\epsilon
\end{equation}
where $\dim V$ is the cardinality of the basis of $V$.

Note that this Betti number depends on the choice of the grouping scale $\epsilon$.

Another way to calculate the Betti number $\beta_k^\epsilon$ is by defining the combinatorial laplacian $\Delta_k^\epsilon$ \cite{berry2022quantifying,towards2022}:
\begin{equation}
    \Delta_k^\epsilon={(\partial_k^\epsilon)}^\dagger\partial_k^\epsilon+\partial_{k+1}^\epsilon{(\partial_{k+1}^\epsilon)}^\dagger
\end{equation}
and getting $\beta_k^\epsilon$ as:
\begin{equation}
    \beta_k^\epsilon=\dim\ker\Delta_k^\epsilon
\end{equation}
Thus, $\beta_k^\epsilon$ is the number of zero eigenvalues of $\Delta_k^\epsilon$ \cite{towards2022}.

\section{Quantum Topological Data Analysis}
\label{qtda}
In 2016, Lloyd et al. \cite{Lloyd2016} proposed the first algorithm to calculate Betti numbers using a quantum computer known as the LGZ (Lloyd, Garnerone and Zanardi) algorithm.
Since then, various improvements and variations have been proposed \cite{georgesiopsis2019,mcardle2022streamlined,ubaru2021quantum,towards2022,akhalwaya2022quantum,quantum2022}. Here, we will use the QPE algorithm to estimate the number of zero eigenvalues of the combinatorial laplacian. We also discuss practical considerations while implementing the algorithm.

Given the grouping scale $\epsilon$ and an integer $0\le k\le n-1$, we start with the combinatorial laplacian $\Delta_k^\epsilon$. Due to its form, the combinatorial laplacian is a real symmetric matrix with dimension equal to the number of $k$-simplicies in $S_k^\epsilon = \lvert{S_k^\epsilon}\rvert$. If $\lambda_j$ is an eigenvalue of $\Delta_k^\epsilon$, then $e^{i\lambda_j}$ is an eigenvalue of the unitary $e^{i\Delta_k^\epsilon}$. The QPE algorithm estimates the eigenvalue of an eigenvector of a unitary operator, i.e. given a unitary matrix $U$ with a quantum state $\ket{\psi}$ such that $U\ket{\psi}=e^{2\pi i\theta}\ket{\psi}$, QPE estimates the value of $\theta$. Thus, $\theta=0$ corresponds to $\lambda=0$ for $\Delta_k^\epsilon$.

The unitary $U$'s shape needs to be $2^q\times2^q$ to act on $q$ qubits. Thus, the matrix $\Delta_k^\epsilon$ also needs to be padded to get the dimension to the nearest power of 2.
While it has been suggested to pad the matrix with all zeros \cite{towards2022}, this has the side effect of increasing the number of zero eigenvalues, thus changing the calculated Betti number, which needs to be corrected post-estimation. Thus, we suggest padding the combinatorial laplacian $\Delta_k^\epsilon$ with an identity matrix with $\Tilde{\lambda}_{max}/2$ in place of ones. Here, $\Tilde{\lambda}_{max}$ is the estimate of the maximum eigenvalue of $\Delta_k^\epsilon$ using the Gershgorin circle theorem \cite{circle} such that:
\begin{align}
    \Tilde{\Delta_k^\epsilon}&=
    \begin{bmatrix}
\Delta_k^\epsilon & 0 \\
0 & \frac{\Tilde{\lambda}_{max}}{2}\cdot I_{2^q-\lvert{S_k^\epsilon}\rvert}
\end{bmatrix}_{2^q\times 2^q}
\end{align}
where $\Tilde{\Delta_k^\epsilon}$ is the padded combinatorial laplacian and $q=\ceil{\log_2\lvert{S_k^\epsilon}\rvert}$ is the number of qubits this operator will act on.

In QPE, as $2\pi\theta$ increases beyond $2\pi$, the eigenvalues will start repeating due to their periodic form. Thus, $\theta$ is restricted to $[0,1)$. As $\lambda\rightarrow2\pi\theta$ this means that $\lambda\in[0,2\pi)$. Thus, we need to restrict the eigenvalues of the combinatorial laplacian to this range. This can be achieved by rescaling $\Tilde{\Delta_k^\epsilon}$ by $\delta/\Tilde{\lambda}_{max}$ where $\delta$ is slightly less than $2\pi$. Thus, the final unitary for the QPE algorithm is:
\begin{align}
    U^\epsilon&=e^{iH^\epsilon}\\
    H^\epsilon&=\frac{\delta}{\Tilde{\lambda}_{max}}\Tilde{\Delta_k^\epsilon}
\end{align}

\begin{figure}
    \centering
    \hspace{-25mm}\resizebox{!}{!}{\input{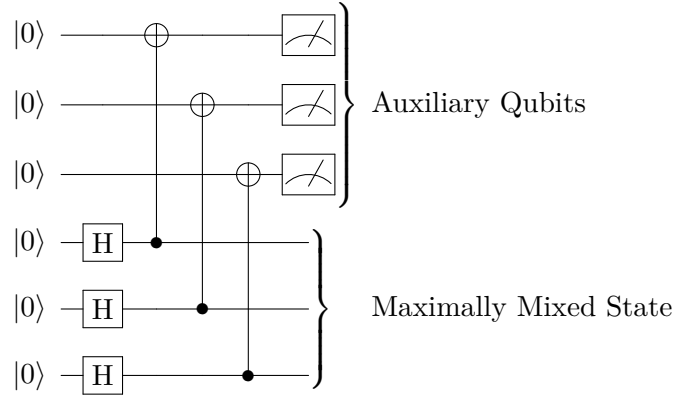}}
    \caption{Circuit creating a 3 qubit maximally mixed state $I/2^3$ using 3 auxiliary qubits.}
    \label{fig:mixed}
\end{figure}

The initial state on which QPE acts is an eigenstate of the unitary, and the $\theta$ value is the corresponding phase of the eigenvalue. Suppose instead; we use the maximally mixed state $I/2^q$ (see Fig.~\ref{fig:mixed}) as the initial state and run the algorithm for $\alpha$ times. In that case, the probability of getting zero in the eigenvalue register is given by:
\begin{alignat}{2}
    &p(0)&&= \frac{\lvert\{i\mid\Tilde{\theta}_i=0\}\rvert}{\alpha}=\frac{\Tilde{\beta}_k^\epsilon}{2^q}\\
    \implies&\Tilde{\beta}_k^\epsilon&&=2^q\cdot p(0)
\end{alignat}
where $\Tilde{\theta}_i$ is the $i^{th}$ estimate of $\theta$ from QPE and $\Tilde{\beta}_k^\epsilon$ is our estimate of the Betti number $\beta_k^\epsilon$.
\paragraph{Note}
As calculated, $\Tilde{\beta}_k^\epsilon$ is a rational number and can be rounded to the nearest whole number depending on the application. These numbers can also be fed directly for further processing, for example, in ML tasks.

\section{Experimental Results}
\label{experiment}
In this section, we present the results of performing the QTDA algorithm on various randomly generated simplicial complexes using a range of shots and precision qubits for QPE.

Starting from a randomly generated simplicial complex for a given $n$, the Betti numbers and combinatorial laplacians are calculated classically for various $k$ values.
$\Tilde{\beta}_k^\epsilon$ is estimated from $\Tilde{\Delta_k^\epsilon}$ using the quantum algorithm.

\begin{figure}
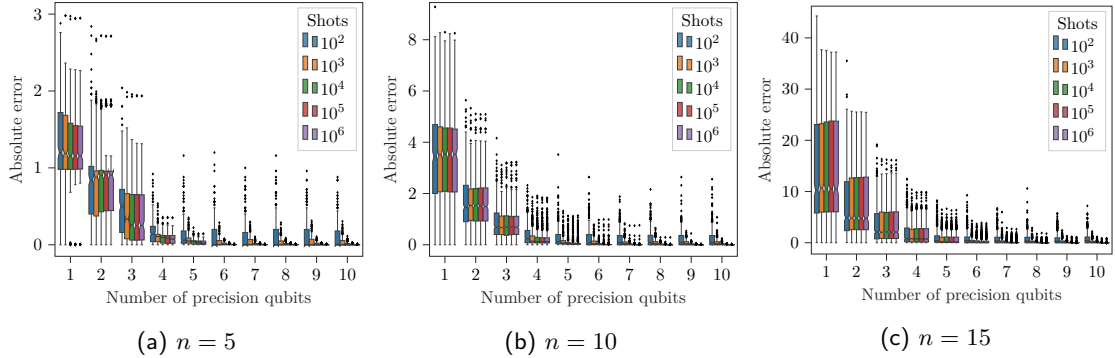

    \centering
    \begin{subfigure}{.32\columnwidth}
        \centering
        \resizebox{1\linewidth}{!}{\input{plots/5_no_round.tex}}
        \caption{$n=5$}
        \label{fig:n5}
    \end{subfigure}
    \begin{subfigure}{.32\columnwidth}
        \centering
        \resizebox{1\linewidth}{!}{\input{plots/10_no_round.tex}}
        \caption{$n=10$}
        \label{fig:n10}
    \end{subfigure}
    \begin{subfigure}{.32\columnwidth}
        \centering
        \resizebox{1\linewidth}{!}{\input{plots/15_no_round.tex}}
        \caption{$n=15$}
        \label{fig:n15}
    \end{subfigure}
    \caption{Boxplots of the absolute error for (\subref{fig:n5}) $n=5$, (\subref{fig:n10}) $n=10$ and (\subref{fig:n15}) $n=15$ using different number of shots and precision qubits. $100$ random simplicial complexes are considered for each $n$.}
    \label{fig:betti_results}
\end{figure}

The absolute error (AE) is defined as:
\begin{equation}
    \text{AE}=\lvert \Tilde{\beta}_k^\epsilon-\beta_k^\epsilon \rvert
\end{equation}
and is shown in Fig.~\ref{fig:betti_results} for $n=\{5,10,15\}$. Increasing the number of precision qubits and/or shots reduces the estimate's error. We also note that the error reduces to zero, given enough resources. The simulations were run using PennyLane \cite{bergholm2018pennylane}.

\section{Classification using QTDA}
\label{classification}

\begin{table}[b]
\centering
\caption{Classification accuracy values for the gearbox features dataset.}
\label{tab:res_gear}
\begin{tabular}{@{}cccc@{}}
\toprule
Precision qubits & Training accuracy & Validation accuracy & Mean absolute error \\ \midrule
1                & 0.980             & 0.892               & 1.121               \\
2                & 0.980             & 0.912               & 0.852               \\
3                & 0.980             & 0.907               & 0.605               \\
4                & 0.980             & 0.902               & 0.142               \\
5                & 0.980             & 0.902               & 0.097               \\ \bottomrule
\end{tabular}
\end{table}

We use the QTDA algorithm to analyse time series data of a gearbox and classify the data as faulty or healthy. This data comes from Southeast University, China is available on GitHub \footnote{https://github.com/cathysiyu/Mechanical-datasets}.

Starting from the time series data for healthy and surface fault gearbox data, data samples are created by taking 500 time stamps at a time. An equal number of random samples are taken from both sets to create the considered dataset. A point cloud is created from the time series using TakensEmbedding from giotto-tda \cite{tauzin2020giottotda}, and a simplicial complex is created using GUDHI \cite{gudhi:RipsComplex}. Betti numbers are estimated similarly from the simplicial complex as in Section~\ref{experiment}. The estimated Betti numbers \{$\Tilde{\beta}_0^\epsilon,\Tilde{\beta}_1^\epsilon\}$ form the new feature set. Classification is performed using scikit-learn \cite{scikit-learn}. Classification accuracy of $100\%$ on the validation set was achieved in this example.

We have also used the processed gearbox data obtained by extracting six features from the time series \cite{gearbox}. This data has a total of 255 data points out of which 51 are healthy and the rest are faulty. Four points in a 3D space are generated for each six-dimensional data point by taking three features at a time. This creates a point cloud for each original data point. QTDA algorithm is applied to this data to estimate the Betti numbers \{$\Tilde{\beta}_0^\epsilon,\Tilde{\beta}_1^\epsilon\}$ which are used in classical logistic regression for classification. The train-validation split used was 20\%-80\%. Table~\ref{tab:res_gear} shows the accuracy values obtained through this process while varying the number of precision qubits for QPE. The accuracy values when using the actual Betti numbers were 0.980 and 0.902 for training and validation sets, respectively. The number of shots in the experiment was fixed at 100. The table also shows the mean absolute error between the estimated and actual Betti numbers for each precision qubit value. The error decreases as the number of precision qubits increases.

\begin{figure}
    \centering
    \resizebox{.4\columnwidth}{!}{\input{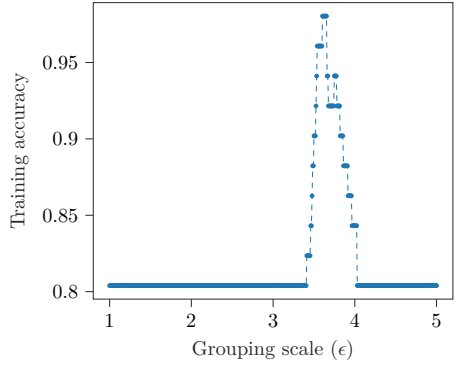}}
    \caption{Plot of training accuracy (using actual Betti numbers) vs the grouping scale ($\epsilon$).}
    \label{fig:group}
\end{figure}

The grouping scale for the first case was a fixed value arrived at using trial and error. It was used as a hyperparameter in the second case, where we repeated the experiment for the training data 50 times for linearly spaced values of $\epsilon\in[3,5]$ to arrive at the optimal grouping scale. Fig.~\ref{fig:group} shows the training accuracy (using actual Betti numbers) vs the grouping scale values. This graph is similar in shape when using estimated Betti numbers.

\section{Conclusion and Discussion}
\label{conclusion}
We have demonstrated the use of QTDA algorithm based on QPE on the combinatorial laplacian to estimate Betti numbers and shown how these can be used for machine learning with time series data suitable for machine diagnostics.
We have also given examples to show how shots and number of precision qubits can affect the estimation error. Depending on the application, very high precision might not be required.

While the extracted features were used with classical ML models, they can also be used with QML models. We can also envision the QTDA algorithm as a sub-circuit of a bigger QML model. Previously, TDA has been used for classification by fusing conventional features with TDA features \cite{eeg}; similar things can be attempted in a QML pipeline.

While this article focuses on calculating Betti numbers which are dependent on the grouping scale, there are algorithms that estimate persistent Betti numbers which are invariant. Those might act as better features for specific scenarios, including when there is noise in the data. This will be part of our future work.
The experiments in this paper were simulated on ideal devices without noise. It will be interesting to see how the algorithm behaves on NISQ devices and how it can be made robust to noise such that the results remain usable. We are also looking at ways to decrease the circuit depth for the algorithm.
\printbibliography

@Article{Lloyd2016,
author={Lloyd, Seth
and Garnerone, Silvano
and Zanardi, Paolo},
title={Quantum algorithms for topological and geometric analysis of data},
journal={Nature Communications},
year={2016},
month={1},
day={25},
volume={7},
number={1},
pages={10138},
issn={2041-1723},
doi={10.1038/ncomms10138},
url={https://doi.org/10.1038/ncomms10138}
}

@Article{circle,
author={S.~Ger{\v s}gorin},
title={\"Uber die Abgrenzung der Eigenwerte einer Matrix},
journal={Bulletin de l'Acad\'emie des Sciences de l'URSS. Classe des sciences math\'ematiques et na},
year={1931},
pages={749--754},
volume={6},
nolink={}
}

@misc{bergholm2018pennylane,
    title={PennyLane: Automatic differentiation of hybrid quantum-classical computations},
    author={Ville Bergholm and Josh Izaac and Maria Schuld and Christian Gogolin and Shahnawaz Ahmed and Vishnu Ajith and M. Sohaib Alam and Guillermo Alonso-Linaje and B. AkashNarayanan and Ali Asadi and Juan Miguel Arrazola and Utkarsh Azad and Sam Banning and Carsten Blank and Thomas R Bromley and Benjamin A. Cordier and Jack Ceroni and Alain Delgado and Olivia Di Matteo and Amintor Dusko and Tanya Garg and Diego Guala and Anthony Hayes and Ryan Hill and Aroosa Ijaz and Theodor Isacsson and David Ittah and Soran Jahangiri and Prateek Jain and Edward Jiang and Ankit Khandelwal and Korbinian Kottmann and Robert A. Lang and Christina Lee and Thomas Loke and Angus Lowe and Keri McKiernan and Johannes Jakob Meyer and J. A. Montañez-Barrera and Romain Moyard and Zeyue Niu and Lee James O'Riordan and Steven Oud and Ashish Panigrahi and Chae-Yeun Park and Daniel Polatajko and Nicolás Quesada and Chase Roberts and Nahum Sá and Isidor Schoch and Borun Shi and Shuli Shu and Sukin Sim and Arshpreet Singh and Ingrid Strandberg and Jay Soni and Antal Száva and Slimane Thabet and Rodrigo A. Vargas-Hernández and Trevor Vincent and Nicola Vitucci and Maurice Weber and David Wierichs and Roeland Wiersema and Moritz Willmann and Vincent Wong and Shaoming Zhang and Nathan Killoran},
    year={2018},
    eprint={1811.04968},
    archivePrefix={arXiv},
    primaryClass={quant-ph},
    doi={10.48550/arXiv.1811.04968}
}

@misc{tauzin2020giottotda,
      title={giotto-tda: A Topological Data Analysis Toolkit for Machine Learning and Data Exploration},
      author={Guillaume Tauzin and Umberto Lupo and Lewis Tunstall and Julian Burella Pérez and Matteo Caorsi and Anibal Medina-Mardones and Alberto Dassatti and Kathryn Hess},
      year={2020},
      eprint={2004.02551},
      archivePrefix={arXiv},
      primaryClass={cs.LG}
}

@incollection{gudhi:RipsComplex
, author    = {Cl{\'{e}}ment Maria and Pawel Dlotko and Vincent Rouvreau and Marc Glisse}
, title     = {Rips complex}
, publisher = {GUDHI Editorial Board}
, edition   = {3.7.1}
, booktitle = {GUDHI User and Reference Manual}
, url       = {https://gudhi.inria.fr/doc/3.7.1/group__rips__complex.html}
, year      = {2023}
}

@article{scikit-learn,
 title={Scikit-learn: Machine Learning in {P}ython},
 author={Pedregosa, F. and Varoquaux, G. and Gramfort, A. and Michel, V.
         and Thirion, B. and Grisel, O. and Blondel, M. and Prettenhofer, P.
         and Weiss, R. and Dubourg, V. and Vanderplas, J. and Passos, A. and
         Cournapeau, D. and Brucher, M. and Perrot, M. and Duchesnay, E.},
 journal={Journal of Machine Learning Research},
 volume={12},
 pages={2825--2830},
 year={2011},
 number={85},
 url={http://jmlr.org/papers/v12/pedregosa11a.html}
}

@INPROCEEDINGS{gearbox,
  author={Kumar, Kriti and Sahu, Saurabh and Majumdar, Angshul and Chandra, M Girish},
  booktitle={2021 International Joint Conference on Neural Networks (IJCNN)}, 
  title={AutoFuse: A Semi-supervised Autoencoder based Multi-Sensor Fusion Framework}, 
  year={2021},
  volume={},
  number={},
  pages={1-7},
  doi={10.1109/IJCNN52387.2021.9533389}
}

@ARTICLE {georgesiopsis2019,
    author   = "George Siopsis",
    title    = "Quantum topological data analysis with continuous variables",
    journal  = "Foundations of Data Science",
    year     = "2019",
    volume   = "1",
    number   = "4",
    pages    = "419--431",
    doi = {10.3934/fods.2019017}
}

@misc{berry2022quantifying,
    title={Quantifying Quantum Advantage in Topological Data Analysis},
    author={Dominic W. Berry and Yuan Su and Casper Gyurik and Robbie King and Joao Basso and Alexander Del Toro Barba and Abhishek Rajput and Nathan Wiebe and Vedran Dunjko and Ryan Babbush},
    year={2022},
    eprint={2209.13581},
    archivePrefix={arXiv},
    primaryClass={quant-ph}
}

@misc{mcardle2022streamlined,
    title={A streamlined quantum algorithm for topological data analysis with exponentially fewer qubits},
    author={Sam McArdle and András Gilyén and Mario Berta},
    year={2022},
    eprint={2209.12887},
    archivePrefix={arXiv},
    primaryClass={quant-ph}
}

@misc{ubaru2021quantum,
    title={Quantum Topological Data Analysis with Linear Depth and Exponential Speedup},
    author={Shashanka Ubaru and Ismail Yunus Akhalwaya and Mark S. Squillante and Kenneth L. Clarkson and Lior Horesh},
    year={2021},
    eprint={2108.02811},
    archivePrefix={arXiv},
    primaryClass={quant-ph}
}

@article{towards2022,
   title={Towards quantum advantage via topological data analysis},
   volume={6},
   ISSN={2521-327X},
   url={http://dx.doi.org/10.22331/q-2022-11-10-855},
   DOI={10.22331/q-2022-11-10-855},
   journal={Quantum},
   publisher={Verein zur Forderung des Open Access Publizierens in den Quantenwissenschaften},
   author={Gyurik, Casper and Cade, Chris and Dunjko, Vedran},
   year={2022},
   month={11},
   pages={855} }

@misc{akhalwaya2022quantum,
    title={Towards Quantum Advantage on Noisy Quantum Computers},
    author={Ismail Yunus Akhalwaya and Shashanka Ubaru and Kenneth L. Clarkson and Mark S. Squillante and Vishnu Jejjala and Yang-Hui He and Kugendran Naidoo and Vasileios Kalantzis and Lior Horesh},
    year={2022},
    eprint={2209.09371},
    archivePrefix={arXiv},
    primaryClass={quant-ph}
}

@article{quantum2022,
   title={Quantum algorithm for persistent Betti numbers and topological data analysis},
   volume={6},
   ISSN={2521-327X},
   url={http://dx.doi.org/10.22331/q-2022-12-07-873},
   DOI={10.22331/q-2022-12-07-873},
   journal={Quantum},
   publisher={Verein zur Forderung des Open Access Publizierens in den Quantenwissenschaften},
   author={Hayakawa, Ryu},
   year={2022},
   month={12},
   pages={873} }

@INPROCEEDINGS{eeg,
  author={Das, Arup Kumar and Kumar, Kriti and Gavas, Rahul D. and Jaiswal, Dibyanshu and Chatterjee, Debatri and Ramakrishnan, Ramesh Kumar and Chandra, M Girish and Pal, Arpan},
  booktitle={2020 28th European Signal Processing Conference (EUSIPCO)}, 
  title={Cognitive Fatigue Detection from EEG Signals using Topological Signal Processing}, 
  year={2021},
  volume={},
  number={},
  pages={1313-1317},
  doi={10.23919/Eusipco47968.2020.9287418}}

\appendix
\section{A worked-out example}
\label{example}
\begin{figure}[b]
    \centering
    \begin{subfigure}{.3\columnwidth}
        \centering
        \includegraphics[width=\linewidth]{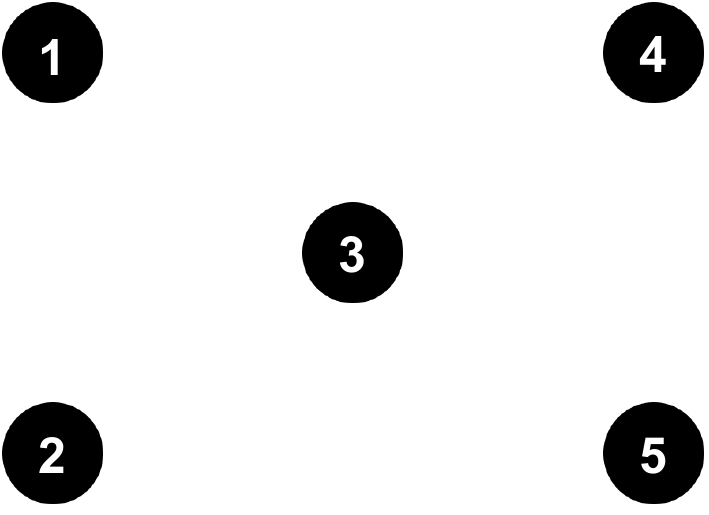}
        \caption{Point cloud}
        \label{fig:eg}
    \end{subfigure}
    \hspace{10pt}
    \begin{subfigure}{.3\columnwidth}
        \centering
        \includegraphics[width=\linewidth]{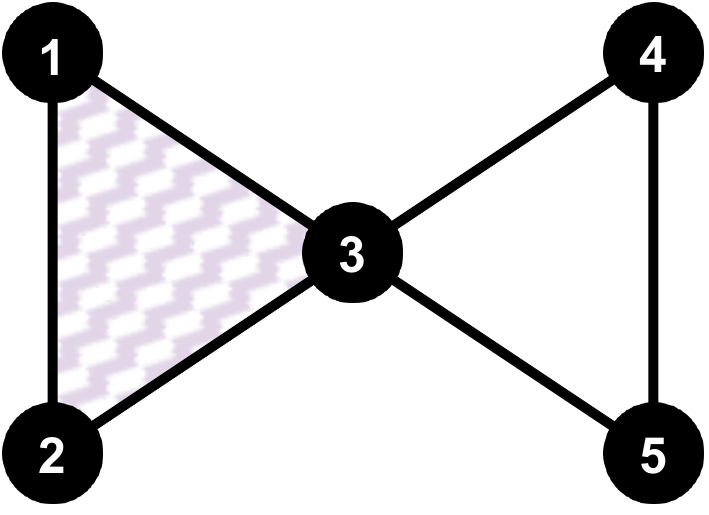}
        \caption{Simplicial complex $\mathcal{K}_\epsilon$}
        \label{fig:eg2}
    \end{subfigure}
    \caption{(\subref{fig:eg}) shows the point cloud used in the example and (\subref{fig:eg2}) shows the simplicial complex formed with the grouping scale $\epsilon$.}
\end{figure}

In this appendix, we demonstrate the QTDA algorithm using an example. Suppose we have a point cloud as shown in Fig.~\ref{fig:eg}. Given a grouping scale $\epsilon$, we can get to the simplicial complex $\mathcal{K}_\epsilon$, shown in Fig.~\ref{fig:eg2}.

\begin{equation}
    \mathcal{K}_\epsilon=\{ \{1\},\{2\},\{3\},\{4\},\{5\},\{1,2\},\{1,3\},\{2,3\},\{1,2,3\},\{3,4\},\{3,5\},\{4,5\} \}
\end{equation}

To calculate $\beta_1^{\epsilon}$ we first need the combinatorial laplacian $\Delta_1^{\epsilon}$. Let's start by calculating the restricted boundary operators $\partial_1^{\epsilon}$ and $\partial_2^{\epsilon}$.

From the simplicial complex, we get,

\begin{align}
    \partial_1^{\epsilon}&=\begin{bmatrix}
1 & 1 & 0 & 0 & 0 & 0 \\
-1 & 0 & 1 & 0 & 0 & 0 \\
0 & -1 & -1 & 1 & 1 & 0 \\
0 & 0 & 0 & -1 & 0 & 1 \\
0 & 0 & 0 & 0 & -1 & -1 
\end{bmatrix}\\
\partial_2^{\epsilon}&=
\begin{bmatrix}
1 \\
-1 \\
1 \\
0 \\
0 \\
0 
\end{bmatrix}
\end{align}

Thus, we get the combinatorial laplacian as,
\begin{align}
    \Delta_1^\epsilon&={(\partial_1^\epsilon)}^\dagger\partial_1^\epsilon+\partial_{2}^\epsilon{(\partial_{2}^\epsilon)}^\dagger\\
    &=\begin{bmatrix}
3 & 0 & 0 & 0 & 0 & 0 \\
0 & 3 & 0 & -1 & -1 & 0 \\
0 & 0 & 3 & -1 & -1 & 0 \\
0 & -1 & -1 & 2 & 1 & -1 \\
0 & -1 & -1 & 1 & 2 & 1 \\
0 & 0 & 0 & -1 & 1 & 2 
\end{bmatrix}
\end{align}

Now, as the size of the matrix is $6\times6$, it needs to be padded to the nearest power of 2, which is 8 $(q=3)$ here.

Thus, the padded combinatorial laplacian is
\begin{equation}
      \Tilde{\Delta_1^\epsilon}=\begin{bmatrix}
3 & 0 & 0 & 0 & 0 & 0 & 0 & 0 \\
0 & 3 & 0 & -1 & -1 & 0 & 0 & 0 \\
0 & 0 & 3 & -1 & -1 & 0 & 0 & 0 \\
0 & -1 & -1 & 2 & 1 & -1 & 0 & 0 \\
0 & -1 & -1 & 1 & 2 & 1 & 0 & 0 \\
0 & 0 & 0 & -1 & 1 & 2 & 0 & 0 \\
0 & 0 & 0 & 0 & 0 & 0 & 3 & 0 \\
0 & 0 & 0 & 0 & 0 & 0 & 0 & 3 
\end{bmatrix}  
\end{equation}
with $\Tilde{\lambda}_{max}=6$. Taking $\delta=6$ we get $H^\epsilon=  \Tilde{\Delta_1^\epsilon}$.

The next step is to get the Pauli decomposition of this Hamiltonian. The Pauli decomposition for $H^\epsilon$ is given by
\begin{alignat}{10}
    H^\epsilon=  &-0.5 ~&&\text{X} \text{X} \text{I}
&&-0.5 ~&&\text{Y} \text{Y} \text{I}
&&-0.5 ~&&\text{Z} \text{I} \text{X}
&&-0.25 ~&&\text{I} \text{X} \text{I}
&&-0.25 ~&&\text{X} \text{I} \text{X} \nonumber\\
&-0.25 ~&&\text{X} \text{Y} \text{Y}
&&-0.25 ~&&\text{X} \text{Z} \text{X}
&&-0.25 ~&&\text{Y} \text{I} \text{Y}
&&-0.25 ~&&\text{Y} \text{Z} \text{Y}
&&-0.25 ~&&\text{Z} \text{X} \text{I} \nonumber\\
&-0.125~&&\text{I} \text{Z} \text{I}
&&-0.125 ~&&\text{I} \text{Z} \text{Z}
&&-0.125 ~&&\text{Z} \text{Z} \text{Z}
&&+0.125 ~&&\text{I} \text{I} \text{Z}
&&+ 0.125 ~&&\text{Z} \text{I} \text{I} \nonumber\\
&+ 0.125 ~&&\text{Z} \text{I} \text{Z}
&&+ 0.25 ~&&\text{I} \text{X} \text{Z}
&&+ 0.25 ~&&\text{X} \text{X} \text{X}
&&+ 0.25 ~&&\text{Y} \text{X} \text{Y}
&&+ 0.25 ~&&\text{Y} \text{Y} \text{X} \nonumber\\
&+ 0.25 ~&&\text{Z} \text{X} \text{Z}
&&+ 0.375 ~&&\text{Z} \text{Z} \text{I}
&&+ 0.5 ~&&\text{I} \text{Z} \text{X}
&&+ 2.625 ~&&\text{I} \text{I} \text{I}
&& &&
\end{alignat}
where, $\{\text{I},\text{X},\text{Y},\text{Z}\}]$ are the Pauli gates.

\begin{figure}[t]
    \centering
    \resizebox{!}{!}{\input{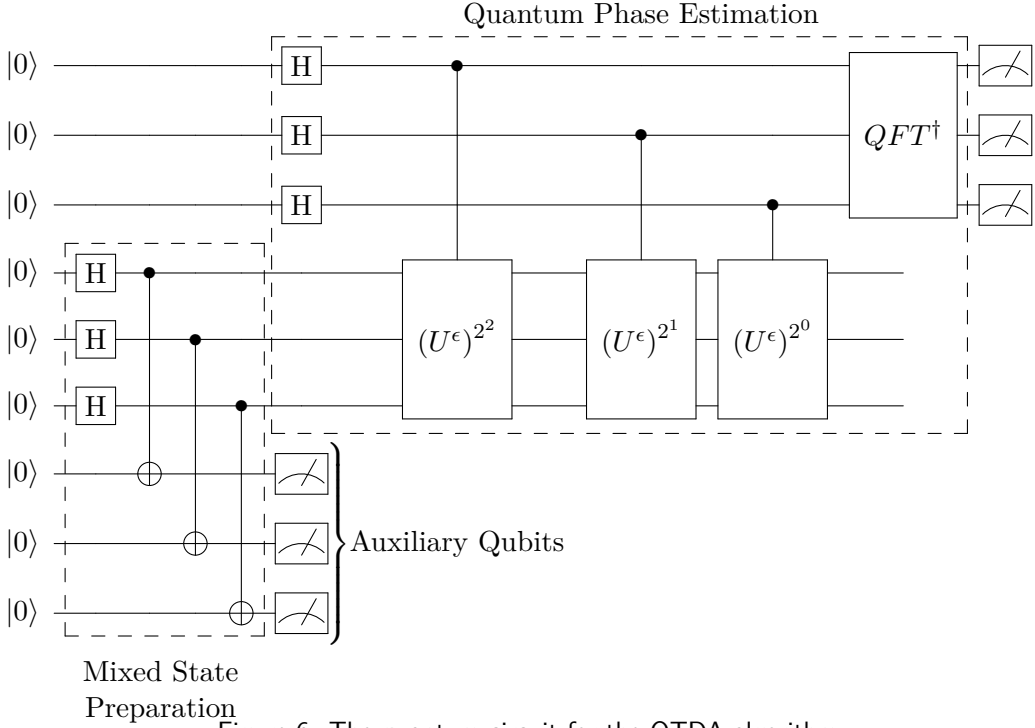}}
    \caption{The quantum circuit for the QTDA algorithm.}
    \label{fig:circ_qtda}
\end{figure}

\begin{figure}
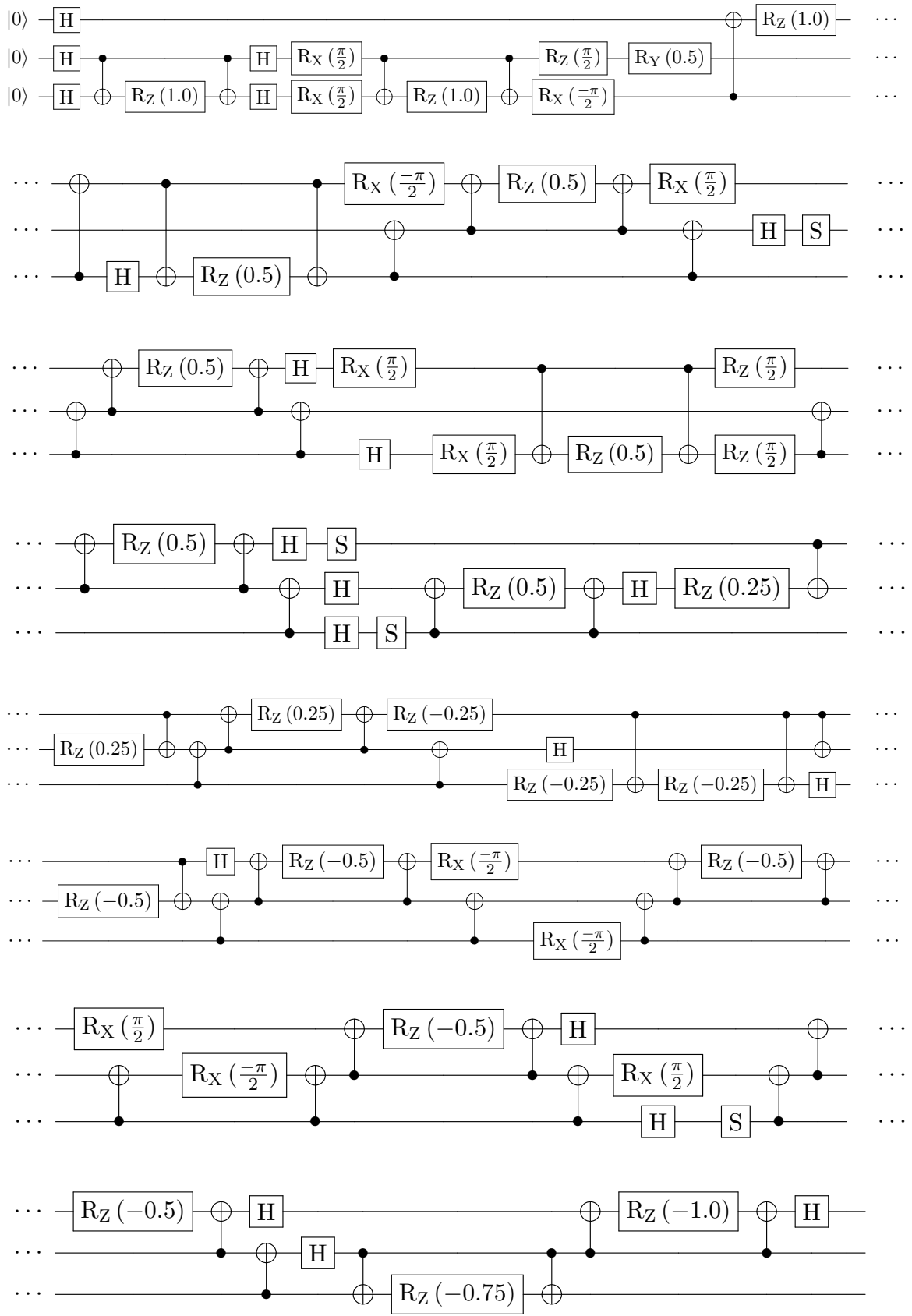

    \centering
    \begin{subfigure}{\columnwidth}
        \resizebox{\columnwidth}{!}{\input{plots/c1.tex}}
    \end{subfigure}
    \begin{subfigure}{\columnwidth}
        \resizebox{\columnwidth}{!}{\input{plots/c2.tex}}
    \end{subfigure}
    \begin{subfigure}{\columnwidth}
        \resizebox{\columnwidth}{!}{\input{plots/c3.tex}}
    \end{subfigure}
    \begin{subfigure}{\columnwidth}
        \resizebox{\columnwidth}{!}{\input{plots/c4.tex}}
    \end{subfigure}
    \begin{subfigure}{\columnwidth}
        \resizebox{\columnwidth}{!}{\input{plots/c5.tex}}
    \end{subfigure}
    \begin{subfigure}{\columnwidth}
        \resizebox{\columnwidth}{!}{\input{plots/c6.tex}}
    \end{subfigure}
    \begin{subfigure}{\columnwidth}
        \resizebox{\columnwidth}{!}{\input{plots/c7.tex}}
    \end{subfigure}
    \begin{subfigure}{\columnwidth}
        \resizebox{\columnwidth}{!}{\input{plots/c8.tex}}
    \end{subfigure}
    \vspace{-4.5pt}
    \caption{The quantum circuit for $U^\epsilon=e^{iH^\epsilon}$. The circuit has a global phase of $\pi/2$.}
    \label{fig:circ_U}
\end{figure}

We can get the circuit for $U^\epsilon=e^{iH^\epsilon}$ from this decomposition. The circuit for the QTDA algorithm with 3 precision qubits is shown in Fig.~\ref{fig:circ_qtda}, and the circuit for $H^\epsilon$ is shown in Fig.~\ref{fig:circ_U}.

After running the circuit for 1000 shots, we get the probability of getting all zero state $p(0)= 0.149 ={\Tilde{\beta}_1^\epsilon}/{2^3}\implies\Tilde{\beta}_1^\epsilon=1.192$. After rounding to the nearest integer, we have $\Tilde{\beta}_1^\epsilon=1$, which is the correct value.

\end{document}